# Strong Modulation of Infrared Light using Graphene Integration with Plasmonic Fano-Resonant Metasurfaces

Nima Dabidian[1], Iskandar Kholmanov[2], Alexander B. Khanikaev[3], Kaya Tatar[1], Simeon Trendafilov[1], S. Hossein Mousavi[1], Carl Magnuson[2], Rodney S. Ruoff[2], Gennady Shvets[1]*

[1]Department of Physics and Center for Nano and Molecular Science and Technology, the University of Texas at Austin, Austin, Texas 78712, United States

[2]Department of Mechanical Engineering and Materials and Science Program, The University of Texas at Austin, Austin, Texas 78712, United States

[3]Department of Physics, Queens College of The City University of New York, Queens, New York 11367, USA and The Graduate Center of The City University of New York, New York, New York 10016, USA

**Abstract:** Plasmonic metasurfaces represent a promising platform for enhancing light-matter interaction. Active control of the optical response of metasurfaces is desirable for applications such as beam-steering, modulators and switches, biochemical sensors, and compact optoelectronic devices. Here we use a plasmonic metasurface with two Fano resonances to enhance the interaction of infrared light with electrically controllable single layer graphene. It is experimentally shown that the narrow spectral width of these resonances, combined with strong light/graphene coupling, enables reflectivity modulation by nearly an order of magnitude leading to a modulation depth as large as 90%. . Numerical simulations demonstrate the possibility of strong active modulation of the phase of the reflected light while keeping the reflectivity nearly constant, thereby paving the way to tunable infrared lensing and beam steering.

## Introduction and motivation

An optical metasurface[1,2] is a two-dimensional (single-layer) metamaterial. Because it is much easier to fabricate than the fully three-dimensional (3D) counterpart, they are already finding applications in a variety of areas such as light manipulation[3-10], biochemical sensing[10-12], nonlinear optics [13,14] and spectrally-selective thermal emission[15] to name just a few. Interesting example for manipulating light using metasurfaces is extremely thin optical components which can mold the wavefront such as optical lenses[16,17], wave-plates[18] and beam steering devices[3,19]. The scalability and universality of the metasurface design enables their deployment across the electromagnetic spectrum. Especially attractive are the metasurfaces designed to operate in the mid-infrared (mid-IR) part of the optical spectrum. That is due to two factors: the variety of technological applications and the limited choice of conventional optical components in the mid-IR spectral region. For example, the $2 \mu m - 20 \mu m$ spectral range hosts important bio-molecular and chemical fingerprints[14] that are being exploited for ultra-sensitive fingerprinting and characterization of molecular monolayers[9,15]. In addition, the atmospheric transparency windows of $3 \mu m - 5 \mu m$ and $8 \mu m - 12 \mu m$ are exploited for a variety of thermal imaging applications[20,21] . These unique properties of infrared radiation are now used to address crucial health, security, and environmental applications [9,22–25], but more rapid progress is impeded by the limited availability of passive and active mid-infrared devices such as sources and detectors [26–30], as well as optical modulators and switches[31–38]. This niche can be potentially filled by active plasmonic metasurfaces.

For example, plasmonic enhancement of light-matter interaction improves the performance and reduces the sizes of detectors[39–41], sensors [42–44] and lasers[45]. When integrated with electrically or mechanically tunable materials or substrates, plasmonic metasurfaces can assume the role of a versatile platform for developing dynamically tunable optical devices, thus paving the way for a variety of technological applications in hyper-spectral imaging[46], single pixel detection[47], and 3D imaging[48], as well as optical modulators and switches [31–38]. Such devices could potentially overcome some of the limitations (e.g., slow response time, shallow modulation depth) of the existing platforms for IR light modulations that utilize liquid crystals[31,32], advanced materials exhibiting metal-insulator phase transitions[33], mechanically stretchable elastomeric materials, or semiconductor interfaces controlled through electrostatic carrier depletion[35,38]. Here we experimentally demonstrate that these limitations can be potentially overcome by integrating spectrally selective plasmonic metasurfaces with single-layer graphene (SLG).

SLG is a promising material for light modulation because its optical conductivity can be tuned by an electrostatic or chemical doping. This property has already been utilized to tune the IR absorption and spectral response of plasmonic metasurfaces[49–53], and it has been shown that the carrier density in SLG can be potentially modulated at tens of MHz rate [51,54,55]. Nevertheless, the modulation depth experimentally demonstrated in mid-IR so far has not exceeded 30% [51] despite some encouraging theoretical predictions[54,56]. The limited effect of SLG on plasmonic metasurfaces in mid-IR is due to two factors: (i) relatively broad plasmonic resonances in that spectral range [57], and (ii) weak absorption of graphene in mid-IR due to the absence of interband transitions [50]. Small modulation depth of the reflected IR radiation produced by electric gating of the SLG puts into question its very viability as an active material for future optical modulators that have much more stringent performance requirements, e.g. tens of decibels (dBs) maximum-to-minimum signal variation.

In this Letter, we experimentally demonstrate that, by integrating an SLG with a high-Q Fano-resonant metasurface, it is possible to modulate mid-IR reflectivity an order of magnitude ($\approx 10 dB$), thus achieving the modulation depth as high as 90% using electrostatic gating. This is accomplished by designing a metasurface that exhibits a spectrally narrow reflectivity peak constrained on the red and blue sides by two distinct Fano resonances. The strong enhancement of the electric field parallel to SLG's surface is shown to result in strong graphene/metasurface coupling that shifts the plasmonic resonances of the metasurface by approximately half of its spectral width. The resulting shift, brought about by the inductive (i.e. essentially loss-free) response of the SLG, is the cause of the deep modulation of the reflected IR light controlled by electrostatic gating. The effects of graphene's strong coupling to the metasurface on the optical response (i.e. the wavelength-dependent reflectivity/phase and the linewidth of the reflectivity peak) are investigated by varying the charged carriers' concentration in the electrically gated SLG.

**Design of the Fano-resonant metasurface**

For the purpose of achieving a spectrally narrow reflectivity peak, we have designed a metasurface which exhibits broadband reflectivity that is greatly reduced at the two nearby wavelengths due to the phenomenon of plasmon-induced electromagnetically induced transparency(EIT) [58,59,4]. The unit cell of the metasurface is shown in Figure 1a, and the SEM image of the fabricated metasurface is shown in Figure 1b. The unit cell consists of three key elements: (i) a metallic wire that electrically connects the neighboring cells in y-direction, and whose functionality is to provide broadband reflectivity of the y-polarized light; (ii) a pair of x-oriented monopole antenna pairs [59,60] attached to the wire, and (iii) a C-shaped antenna (CSA) placed in their proximity. The plasmonic metasurface is assumed to be placed on a thick $SiO_2$ substrate. In reality, the substrate that is actually used in the experiments is a finite-thickness ($t = 1\mu m$) thermal $SiO_2$ grown on a Si wafer, but the oxide layer is thick enough to avoid any effect of the Si wafer on optical properties of the metasurface. This metasurface exhibits an optical response with two maxima in the transmission spectrum (or two deep reflectivity minima) due to Fano resonances, and therefore can be characterized as a plasmonic metasurfaces exhibiting double electromagnetically induced transparency (double-EIT). This effect is illustrated in Figure 1b, where we use a bottom-up approach of adding different constitutive parts of the unit cell to illustrate the nature of the two resonances as detailed below.

First, one recognizes that a grid of plasmonic wires behaves as anisotropic "dilute plasma" [61] for the y-polarized incident light. Specifically, the structure strongly reflects the light with the wavelength $\lambda$ which is longer than the characteristic effective plasmon wavelength $\lambda_p^{eff}$ which is a function of the inter-wire spacing $P_x$, wire width $w$, and the frequency-dependent dielectric permittivity $\epsilon_s(\lambda)$ of the underlying substrate. As expected, the reflectivity $R(\lambda)$ of the wire array shown as the black solid line in Figure 1c increases as the function of the wavelength and reaches 60% at $\lambda = 7.5\mu m$. Next, a CSA supporting a dipolar antenna resonance at $\lambda_d$ is added to the metasurface. The destructive interference between anti-parallel currents in the wires and in the CSA for $\lambda > \lambda_d$ produces a pronounced dip in reflectivity at $\lambda_2 \approx 7.2\mu m$ as shown by the blue line in Figure 1c. We refer to this dip as the EIT2. Although the dip at $\lambda = \lambda_2$ is not particularly narrow because of the strong coupling between the dipole-active mode of the CSA and the plasmonic wire, we demonstrate below that the half width at half maximum (hwhm) $\sigma_{1/2}$ of the emerging reflectivity peak can be considerably narrowed by employing a second Fano resonance at the shorter wavelength.

Specifically, by adding the x-directed (horizontal) plasmonic monopole antennas to the wire grid, we introduce a second dark (monopole) mode whose resonant frequency corresponding to $\lambda_m \approx 6.2\mu m$ is primarily determined[59,60] by the monopoles' length $l_m$ according to $\lambda_m \approx 4\sqrt{\epsilon_s}(l_m)$. This emergent mode is referred to as the monopole mode because of the strong current flow between neighboring monopole antennas. Note that if the monopole antennas are equally spaced, that is $P_y = 2(L_y + w)$, then by symmetry the monopole mode is strictly dark (i.e. completely decoupled from the normally incident light and the concomitant uniform wire current) in the absence of the near-field coupling between the C-shaped and monopole antennas. The symmetry breaking produced by the introduction of the CSA antennas and by the non-equal spacing of the antennas causes the narrow-band monopole mode to couple to the broad-band wire grid currents, thus producing Fano interference. The destructive Fano interference between the wire current and the current flowing between the monopole antennas results in the short-wavelength reflectivity dip at $\lambda_1 \approx 6.2\mu m$ which is referred to as the first EIT (EIT1) in the rest of the paper. The charge distribution and current profile for the EIT1 dip are shown in Figures 1d,e, respectively.

The addition of the monopole antennas also has an effect on the dipole mode and, by extension, on the EIT2 reflectance dip. By comparing the spectral positions of the EIT2 with (red line) and without (blue line) the monopole antennas, we conclude that the capacitive coupling between the CSAs and monopole antennas causes the spectral position of EIT2 to red-shift. The near-field coupling between C-shaped and monopole antennas produce a characteristic quadrupole-like charge distribution shown in Figure 1f which is responsible for electric field enhancement between the antennas. This field enhancement plotted in Figure 1i can be exploited in graphene-functionalized metasurfaces as explained below. The near-field coupling between the two discrete (dipolar and monopolar) modes and the broad-band currents flowing in the plasmonic wire grid cause Fano interference. The result of this interference is a double-dip reflectivity spectrum shown in Figure 1c (red line). Qualitatively, the EIT2 reflectivity dip at $\lambda_2 \approx 7.6 \mu m$ is caused by the counter-flowing currents in the CSA and the wire grid plotted in Figure 1g. The resulting narrow reflectivity peak "sandwiched" between the EIT1,2 dips at $\lambda_1$ and $\lambda_2$ is the main consequence of the double-Fano resonance. It is exploited in this letter in order to develop a graphene-based reflection modulator. The modulation is accomplished by electrically controlling the conductivity of graphene, which in turn detunes the reflectivity dips by a considerable fraction of the linewidth $\sigma_{1/2}$ of the reflectivity peak.

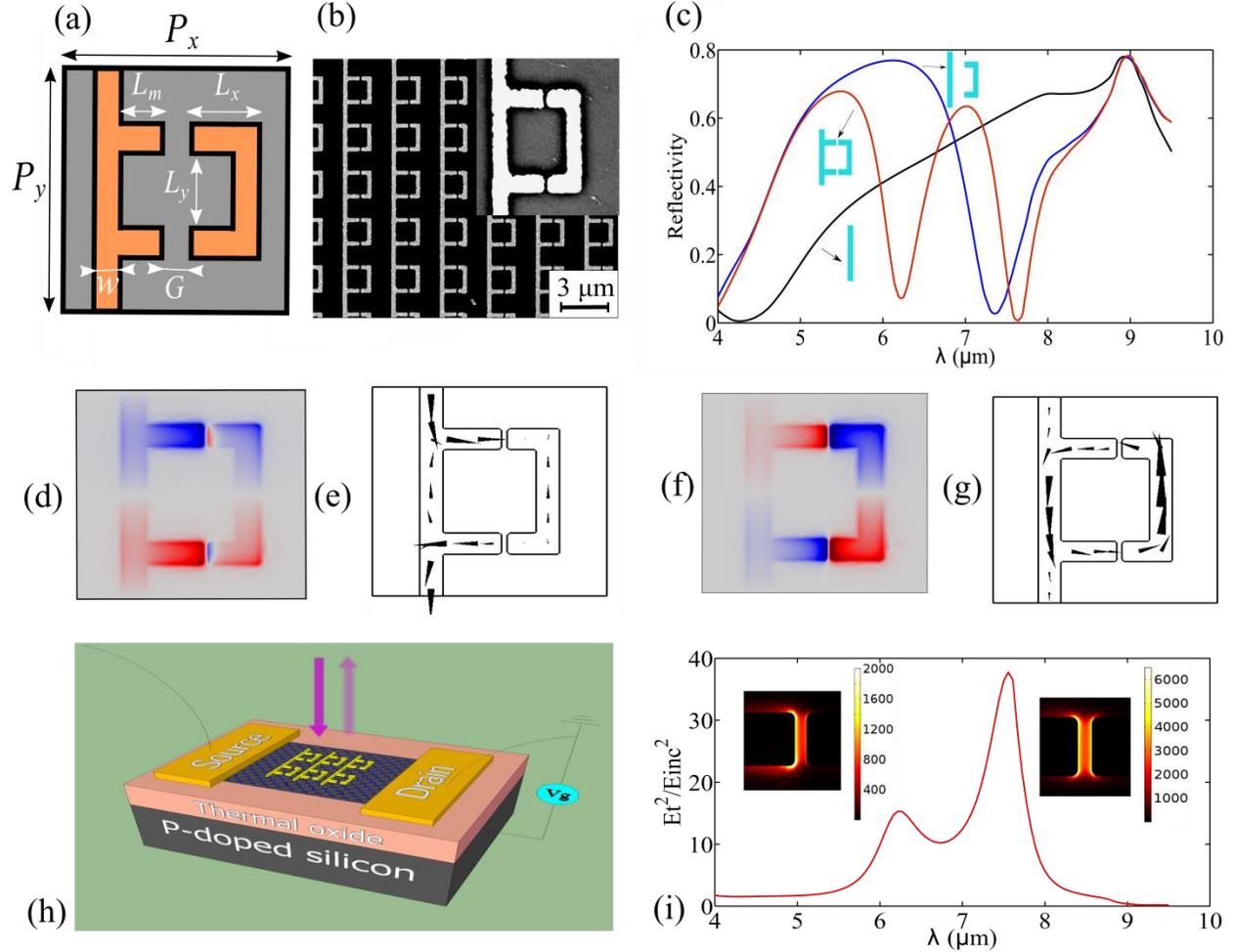

**Figure 1**: **Design of a metasurface exhibiting double-Fano resonance.** (a) Geometry of the unit-cell of the metasurface with parameters: $P_x = P_y = 3\mu m, L_x = .675\mu m, L_m = .75\mu m, L_y = 1.2\mu m,$ and $w = .3\mu m$. G=70nm and 100nm for the two fabricated metasurface. The thickness of metal is 30nm (5nm Cr+25nm gold) (b) SEM picture of the structure on CDV graphene. (c) Simulated Reflectivity at normal incidence for wire grid (black), wire and CSA (blue) and the full structure all on the Si/SiO$_2$ substrate. In (d) and (f) colors represent $E_z$ calculated 5 nm below the metasurfaces/SiO$_2$ interface for EIT1 (d) and EIT2 (f) excited by a y-polarized incident field. The electric current density for EIT1 (e) and EIT2 (g) respectively 5nm above the substrate. (h) Schematics of the sample with the gating geometry: the spacer represents a $t = 1\ \mu m$-thick silicon dioxide. The metasurface is placed on top of graphene. $V_g$ is the back-gate DC voltage. (i) Value of $(E_t/E_{inc})^2$ integrated over graphene surface where $E_t$ is the tangential electric field. The insets show the value of $(E_t/E_{inc})^2$ for the two modes with the colorbar indicating the local values in the gap. All simulations were performed using finite-elements method COMSOL Multiphysics simulation package.

In order to achieve the desired frequency shift of the resonances, it is important that the SLG placed underneath the metasurface experience considerable tangential component $E_t$ of the electric field. Therefore, one of the functions of the metasurface is to produce a strong near-field enhancement $\eta = (E_t/E_{inc})^2$ of the incident electric field $E_{inc}$ on the surface of graphene. This tangential field intensity enhancement, which is proportional to the interaction energy between graphene and the metasurface, is plotted in Figure 1f as the function of the wavelength of the incident wave. The maxima of $\eta$ correspond to the spectral positions of the Fano resonances[9]. Note that a much stronger tangential field enhancement is observed at $\lambda = \lambda_2$ compared to $\lambda = \lambda_1$, suggesting that the strongest coupling between the metasurface and SLG should be expected at the long-wavelength (second) Fano resonance.

We note in passing that SiO$_2$ used as a substrate in this work has an absorption line associated with the excitation of the optical phonons centered at $\lambda_{SiO_2} \approx 9 \mu m$[62,63]. The strongly dispersive character of the substrate's dielectric permittivity $\epsilon_{SiO_2}(\omega) \equiv \epsilon_r + i\epsilon_{im}$ in the frequency range of interest plays a non-negligible role in determining the spectral position and quality factor of both resonances. Specifically, the resonant behavior of $\epsilon_r(\omega)$ causes the quality factor and spectral shift of the dipole mode to increase, compared with the case of a simplified (non-dispersive) substrate that is typical for visible/near-IR frequency range. Additional information about the substrate effect on the quality factor and frequency of the two resonances is provided in the Supporting Information section.

We also note that the possibility of employing two discrete states strongly interfering with a continuum state has been discussed in the original paper by U. Fano[64], and its application to multi-band sensing applications was later suggested [65]. Here, we employ the double-Fano resonance for several purposes. *First,* we use the two Fano resonances in order to produce a narrow-band reflectivity maximum that can be utilized for dynamic modulation of reflectivity using SLG. *Second,* we experimentally demonstrate a dual-band SLG modulator that produces the highest degree of optical modulation at the frequencies corresponding to the two EIT resonances, $\lambda_1$ and $\lambda_2$. *Third,* we demonstrate that the degree of optical modulation (both in amplitude and phase), which is proportional to the frequency shifts $\Delta\omega_{1,2}$ due to the SLG's hybridization with the metasurface, depends on the charge distribution in graphene for each given resonance. For example, the charge distribution on the metasurface corresponding to the second (long-wavelength) Fano resonance supports much higher tangential field, resulting in $\Delta\omega_2 > \Delta\omega_1$. The consequence of this hierarchy of Fano resonances is that the modulation depth around $\lambda = \lambda_2$ is larger than that around $\lambda = \lambda_1$.

**The effect of single-layer graphene on the Fano resonances and metasurface reflectivity**

Extending the earlier developed theory[50,66] of Fano-resonant metasurfaces which demonstrated that the complex reflectivity coefficient r ($\omega$) from a metasurface supporting a single Fano resonance can be described by a two-pole (double-Lorentzian) reflectivity function r ($\omega$) (where the complex frequency poles contain both the spectral positions and the lifetimes of the bright/dark resonances), we approximate the reflectivity from a double-Fano resonant metasurface using the following triple-Lorentzian function:

$$r(\omega) = \frac{A_0}{i\omega + 1/\tau_0} + \frac{A_m}{i(\omega - \omega_m) + 1/\tau_m} + \frac{A_d}{i(\omega - \omega_d) + 1/\tau_d} \quad (1)$$

The first term in eq 1 describes the broadband reflectivity by the plasmonic wires, the second term describes the resonant reflectivity by the monopole resonance, and the third term describes the resonant reflectivity by the CSA resonance. The complex amplitudes $A_0, A_m, A_d$ (where $A_0 \gg A_m, A_d$, see the details in the Supplemental Online Materials section) of the corresponding modes are proportional to their far-field coupling strength [50,66]. Note that, in general, the resonance frequencies do not coincide with the spectral positions of the reflectivity dips: $\omega_{1(2)} \neq \omega_{m(d)}$. The corresponding quality factors of the two resonances that will be computed from the experimental data by fitting the reflectivity spectrum to the $|r(\omega)|^2$ function given by the eq 1 are defined as $Q_{m(d)} = \omega_{m(d)} \tau_{m(d)}$. The convenience of the multi-pole expansion given by eq 1 is in that it enables us to investigate the effect of the SLG on the quality factors and spectral positions of the two Fano resonances. This is done by extracting the frequencies $\omega_{m(d)}$ and lifetimes $\tau_{m(d)}$ of the modes, which depend on the presence/optical properties of the SLG. Specifically, the addition of a SLG perturbs the resonant frequencies of the modes according to $\widetilde{\omega}_{m(d)} = \omega_{m(d)} + \Delta\widetilde{\omega}_{m(d)}$, where the graphene-induced complex-valued frequency shift $\Delta\widetilde{\omega}_{m(d)}$ is calculated [50] according to

$$\Delta\widetilde{\omega}_{m(d)} = \frac{\sigma_{SLG}}{i} \frac{\int_S |E_t|^2 dS}{W_0^{m(d)}}, \quad (2)$$

where $\sigma_{SLG}(\omega) = \sigma_{re} + i\sigma_{im}$ is the complex-valued surface conductivity of the SLG, $E_t$ is the tangential electric field of the mode on the graphene's surface $S$, and $W_0^{m(d)}$ is the stored energy of the given (monopole or dipole) mode. It follows from Eqs. (2, 1.0) that the addition of graphene causes (a) spectral shift of the resonant frequencies $\Re e[\Delta\widetilde{\omega}_{m(d)}]$ proportional to the imaginary (reactive) part $\sigma_{im}$ of graphene's conductivity, and (b) additional spectral broadening of the mode, i.e. the increase of the quantity $1/\tau_{m(d)}$, which is defined by $\Im m[\Delta\widetilde{\omega}_{m(d)}]$ and is proportional the real (resistive) part $\sigma_{re}$ of graphene's conductivity. According to eq 2, these frequency/lifetime changes are proportional to the square of the tangential electric field of each mode on graphene, thus suggesting that the dipole resonance should experience larger perturbation that the monopole resonance according to Figure 1i. Spectral tuning of the optical response of the graphene-functionalized metasurface is accomplished by controlling the conductivity of graphene $\sigma_{SLG}(\omega)$ using electrostatic back-gating as described below.

Graphene's conductivity $\sigma_{SLG}(\omega)$ depends on the areal concentration $n$ of the free carriers (electrons or holes) which is parameterized by the Fermi energy $E_F = \hbar V_F \sqrt{\pi n}$, where $V_F$ is the Fermi velocity. For the typical $V_F = 1.1 * 10^8 \frac{cm}{s}$, the Fermi energy is related to $n$ through a simple relationship $E_F = \sqrt{n[cm^{-2}]}/(7.8 * 10^6)$ eV. The carrier density is controlled by an applied electrostatic potential difference between SLG and the Si backgate according to $n = C_g \Delta V/e$, where $\Delta V = V_g - V_{CNP}$ is the potential deviation from the charge neutrality point (CNP) voltage $V_{CNP}$ that can be experimentally determined from electric measurements [67] as shown below in Figure 2a, and $C_g = \varepsilon/d$ is the gate capacitance per unit area; d and $\varepsilon$ are the thickness and electrostatic permittivity of the SiO$_2$ spacer. We estimate that for our sample $C_g \approx 2nF\ cm^{-2}$. Under the random phase approximation (RPA)[68], graphene's conductivity can be broken up into two contribution according to $\sigma_{SLG}(\omega) = \sigma_{inter}(\omega) + \sigma_{intra}(\omega)$, where

$$\sigma_{inter}(\omega) = \frac{e^2}{4\hbar}\left[\frac{1}{2} + \frac{1}{\pi}\tan^{-1}\left(\frac{\hbar\omega - 2E_F}{2k_BT}\right) - \frac{i}{2\pi}\ln\frac{(\hbar\omega + 2E_F)^2}{(\hbar\omega - 2E_F)^2 + (2k_BT)^2}\right], \quad (3)$$

$$\sigma_{intra}(\omega) = \frac{e^2}{4\hbar}\frac{8ik_BT}{\pi\hbar(\omega + i\tau^{-1})}\ln\left[2\cosh\left(\frac{E_F}{2k_BT}\right)\right],$$

are caused by the interband and intraband transitions, respectively. In eq 3, $k_B$ is the Boltzmann constant, $T$ is the temperature, $\omega$ is the frequency of IR light, and $\tau$ is the carrier collisional time. For this sample $\tau = 18fs$ (corresponding to the carrier scattering rate $\tau^{-1} = 293cm^{-1}$) was estimated using a perturbation theory[50] as detailed in the Supporting Information section.

The real and imaginary parts of the graphene conductivity $\sigma_{SLG}(\omega)$, measured in the units of universal optical conductivity $\sigma_0 = e^2/4\hbar$, are plotted in Figures 2b,c respectively, as functions of the wavelength $\lambda = 2\pi c/\omega$. Note that the real part of $\sigma_{SLG}(\omega)$ is associated with the resistive (lossy) response of the graphene, while its imaginary part represents the reactive response[50]. The plasmonic (i.e. reactive) response of graphene can be quantified by the ratio $(\sigma_i/\sigma_r)$[50]. As Figure 2d shows, the higher the charge concentration, the larger this ratio becomes, making graphene increasingly plasmonic. Below we present experimental results demonstrating that the electrically controllable plasmonic response of graphene can be used to dynamically shift the spectral position of the reflectivity peak shown in Figure1c, thus producing strong modulation.

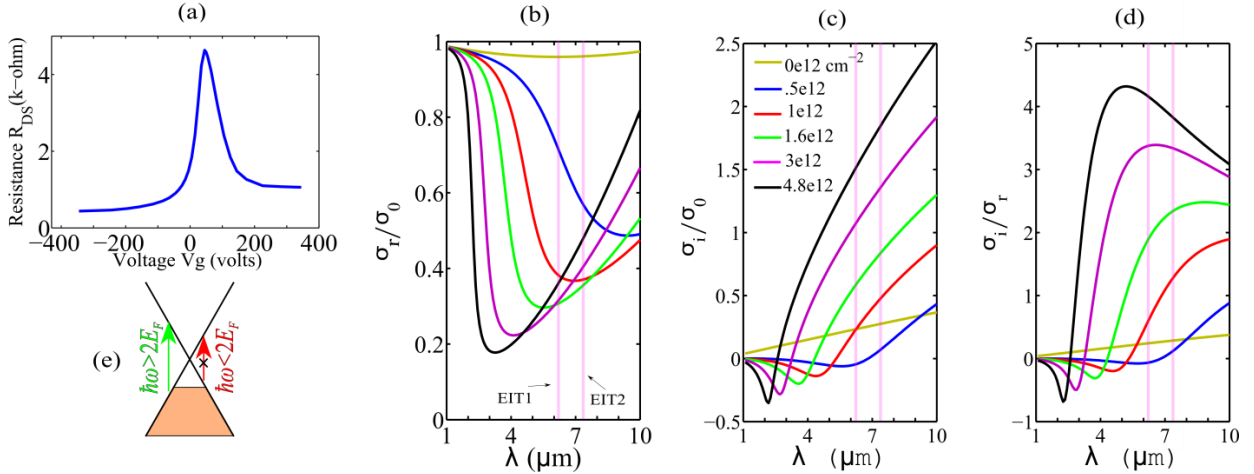

**Figure 2. Electrical and optical properties of graphene**: **(a)** Experimentally obtained dc electric resistance $R_{DS}$ of the SLG measured between the drain and source contacts as a function of the gate voltage $V_g$ (see Figure 1). Charge neutrality point is $V_g \equiv V_{CNP} \approx +45V$. Hole doping of the SLG: $V_g < V_{CNP}$. **(b)** Real and **(c)** imaginary parts of the graphene's optical sheet conductivity $\sigma_{SLG}(\omega)$ as calculated from Eq 3 for different values of graphene doping. **(d)** $\sigma_i/\sigma_r$ for different graphene doping. Pink shaded regions show spectral positions of the two EIT resonances (indicated as EIT1 and EIT2) of the metasurface. Rapid increase of $\sigma_i/\sigma_r$ for high levels of doping is due to increasing plasmonic properties and the decreasing loss owing to Pauli blocking of interband transitions for low-

energy mid-IR photons, as illustrated in (**e**). Hole densities in graphene used in (**b-d**) are color-coded according to the inset in (**c**).

**Fabrication and graphene characterization**

The process of sample fabrication followed the steps below. First, the SLG was grown on polycrystalline Cu foil using a CVD technique [28] and subsequently transferred from the Cu foil onto $1\mu m$ thick insulating (SiO$_2$) layer that was grown on a lightly doped silicon substrate[29] using wet thermal oxidation. Second, two $100\mu m \times 100\mu m$ metasurface samples with unit cell dimensions given in Figure 1 and two different gaps ($G = 70nm$ for Sample 1 and $G = 100nm$ for Sample 2) between the CSAs and monopole antennas were fabricated on top of the SLG using electron beam lithography (EBL). The thickness of the metasurface was 30nm (5nm of Cr and 25nm of Au). An SEM image of a segment of the Sample 1 is shown in Figure 1b, where the inset zooms in on a single unit cell of the metasurface on SLG. Finally, source and drain contacts (10nm Cr+100nm Au) were deposited on top of graphene on both sides of the samples using another EBL step. Back-gating voltage applied across the SiO2 insulating layer between the Si substrate and the drain electrode was used to modulate graphene's carrier density as shown in Figure 1f.

SLG's characterization was carried out using current-voltage (I-V) measurements. The charge neutrality point (CNP) $V_g = V_{CNP} = 45V$ corresponding to $E_F = 0$ is identified by measuring the drain-source electric resistance $R_{DS}(V_g)$ as shown in Figure 2a. The hole mobility $\mu_h$ of graphene at room temperature was calculated from the measured electrical conductivity according to $\mu_h = \sigma(n_h)/n_h e \approx 3800 \ cm^2/Vs$ where $\sigma(n_h)$ is the electrical conductivity of graphene at the hole concentration of $n_h$. We used $n_h = 4.8e12 \ cm^{-2}$ for this calculation. The carrier collisional time can also be calculated[73] from the measured dc electrical conductivity to be $\tau = \frac{\sigma \hbar^2}{2e^2 E_F} \approx 14fs$ which is consistent with the value of $\tau$ derived using the optical conductivity of graphene as explained in the supporting information section.

The slight p-doping of the SLG by the SiO$_2$ substrate is inferred from $V_{CNP} > 0$. Due to the breakdown voltage of silicon dioxide at 0.5 GV/m, we vary the back gate voltage in the $-345V < V_g < 45V$ range using "Heathkit 500V PS-3" power supply. The holes' areal concentration (given by $n_h = C_g \Delta V/e$) can reach the maximum values of $n_h^{max} \approx 4.8e12 \ cm^{-2}$ for the peak gate voltage. According to Figures 2b-d, this variation in SLG's carrier density is expected to change SLG's optical conductivity by almost an order of magnitude.

**Experimental results: deep reflectivity modulation using gated graphene**

Optical characterization of graphene-functionalized back-gated metasurfaces has been performed in the mid-infrared part of the spectrum using a Thermo Scientific Continuum microscope coupled to a Nicolet 6700 FTIR Spectrometer with NA=0.58. The sample spectra averaged over 32 scans were collected with spectral resolution of $4 \ cm^{-1}$. The results of applying gating voltage to the SLG are reported in Figures 3a and 3b (for the Samples 1 and 2, respectively), where the reflectivity spectra from graphene-integrated Fano-resonant metasurfaces are plotted for different carrier (hole) concentrations in the SLG. The reflectivity curves are color-coded according to the values of $n_h$, and the colors are consistent with those used in Figure 2. The interaction strength between graphene and metasurface is determined by the electromagnetic fields in the gap. By choosing a small size for the gap, the metasurface can strongly enhance the tangential electric fields and therefore the electromagnetic interaction with graphene. Coupled with a narrow spectral width of the transmission peak, this strong graphene/metasurface interaction causes strong modulation of the reflected intensity close around $\lambda_2^{mod} = 7\mu m$ and considerably weaker modulation around the monopole mode at $\lambda_1^{mod} = 6\mu m$. Naturally, the strongest modulation is observed close to EIT1 and EIT2 reflectivity dips.

To quantify the efficacy of the resulting graphene/metasurface modulator, we define the wavelength-dependent modulation depth (MD) as [56,68,69]

$$MD(n,\lambda) = \left|\frac{R_{on}(\lambda) - R_{off}(n,\lambda)}{R_{on}(\lambda)}\right| \times 100\% = \left|1 - \frac{1}{RR(n,\lambda)}\right| \times 100\%. \qquad (4)$$

Here $RR(n,\lambda) = R_{on}(\lambda)/R_{off}(n,\lambda)$ is the relative reflectivity that represents the ratio between the baseline reflectivity $R_{on}(\lambda) \equiv R(n = 0, \lambda)$ measured in the absence of free carriers in the SLG and the extinguished reflectivity $R_{off}(n,\lambda) \equiv R(n,\lambda)$ measured for the gate voltage corresponding to the finite gate-induced carrier concentration $n = n_h$. Relative reflectivities $RR(n,\lambda)$ for the Samples 1 and 2 are plotted in Figures 3c,d, respectively. Simply put, the modulation depth represents the percentage of the baseline reflectivity that is extinguished by applying back gate voltage to the SLG. The higher is this extinguishing factor (and, correspondingly, the higher is the relative reflectivity), the better is the modulator. Ideally, one would prefer that both the baseline reflectivity $R_{on}(\lambda)$ and the relative reflectivity $RR(n_h^{max}, \lambda)$ corresponding to the largest applied voltage be large numbers. For example, a modulator with a low value of the baseline reflectivity would be inefficient no matter how high is its relative reflectivity. As our experimental results presented in Figures 3c,d indicate, these two requirements can indeed be satisfied by a Fano-resonant metasurface integrated with back-gated SLG.

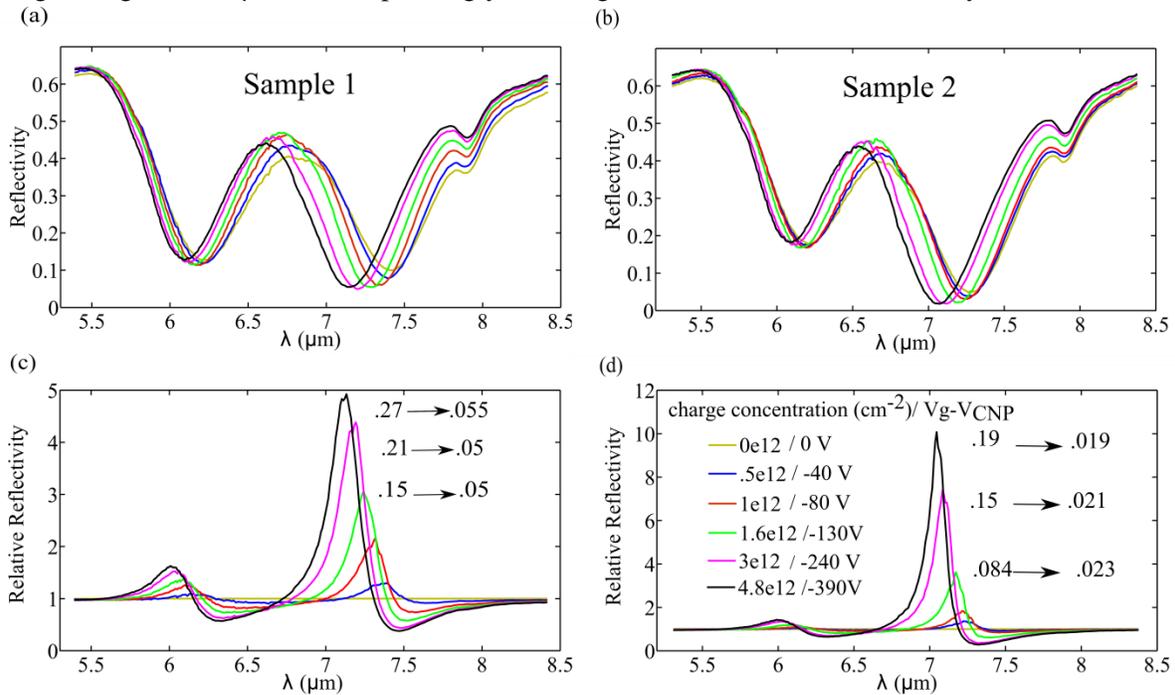

**Figure 3: Experimentally measured reflectivity modulation using back-gated graphene. (a,b)** Reflectivity spectra for different back gate voltages $\Delta V = V_g - V_{CNP}$ for the structures with gap sizes (a) $G = 70nm$ and (b) $G = 100nm$. The spectra are color-coded according to the values of $\Delta V$ consistently with Figure 2. **(c,d)** Relative reflectivities RR$(n,\lambda)$ corresponding to the spectra in (a) and (b). The numbers in the inset represent the baseline reflectivity $R_{on}(\lambda)$ at $\Delta V = 0$ and the extinguished reflectivity $R_{off}(n_h,\lambda)$ for the 3 highest doping levels corresponding to the maximum of RR$(n,\lambda)$.

Specifically, large values of the peak RR, defined as $RR_{peak} \equiv \max_\lambda RR(n_h^{max}, \lambda)$, are experimentally measured around $\lambda_2^{mod}$ for both samples. For Sample 1, $RR_{peak}^{(1)} \approx 5$ leading to $MD_{peak}^{(1)} \approx 80\%$ (corresponding to extinguishing the baseline reflectivity from 27% down to 5.5%) is demonstrated, while an even higher $RR_{peak}^{(2)} \approx 10$ leading to $MD_{peak}^{(2)} \approx 90\%$ (corresponding to extinguishing the baseline reflectivity from 19% down to 1.9%) is achieved for Sample 2. We note that, because of the resonant nature of the metasurface, these modulators are relatively narrow band. For example, the 3dB bandwidth of the modulation depth was measured to be $\Delta\lambda^{(1)} = 245nm$ and $\Delta\lambda^{(2)} = 140nm$ (corresponding to 3.5% and 2% of the resonance wavelength) for Samples 1 and 2, respectively. We also observe that the RR is much smaller for the shorter-wavelength monopole mode at $\lambda_1^{mod} = 6\mu m$. This is a direct consequence of the weaker tangential field enhancement $\eta$ at the monopole resonance as observed from the theoretical calculation shown in Figure 1i.

Note that the small feature at $\lambda \approx 7.9\mu m$ is due to the occurrence of epsilon near zero (ENZ) effect that stems from the longitudinal phonon polariton resonance of $SiO_2$ where $\varepsilon_r^{SiO_2} = 0$ [70]. This resonance is excited by the electric field component of the incident light ($E_z$) which is normal to the substrate. Therefore it is predicted to appear for finite incidence angle and P-polarized incident field. These are indeed the experimental conditions encountered in our experiment because of the inherent properties of the focusing optics of the infrared microscope.

As mentioned earlier, an ideal modulator needs to be both efficient (high baseline reflectivity) and possess a high modulation depth. Achieving high MD by itself is not particularly challenging: as long as the reflectivity drops to near-zero value (as it is indeed the case according to Figures 3a,b of the metasurface under this study), even the slightest spectral detuning of the Fano resonance will result in a high MD. Simultaneously satisfying the efficiency requirement is more challenging because it requires that the spectral detuning $\Delta\lambda_{m(d)}$ due to graphene be comparable to $\sigma_{1/2}$. Therefore, in order to understand how the SLG shifts the frequencies and quality factors of the two Fano resonances, we have fitted the experimentally measured reflectivity spectra $|r(\omega)|^2$ for the Sample 1 to the triple Lorentzian formula given by eq 1. The details of the fitting procedure and the values of the fitting parameters can be found the Supplemental Online Information section. The results of the fitting procedure for the three selected concentrations of charge carriers are shown in Figure 4a. The ENZ mode at $\lambda \approx 7.9\mu m$ has been excluded from the fitting window. The resonant wavelengths $\lambda_{m(d)}$ and quality factors $Q_{m(d)}$ for the two resonances are plotted in Figure 4 for different doping levels. As can be seen from the Figure 4b,c as $n$ increases from zero, the resonant wavelength slightly increases first for small values of $n$ because $\sigma_i(\lambda_{m(d)}) < 0$ due to the interband term [36,50] as shown in Figure 2c.

This spectral detuning is very small when $2E_F(n) < \hbar\omega_{d(m)}$ as marked by the dashed lines in Figures b-e. At the same time, as the Pauli blocking starts taking place for $2E_F(n) > \hbar\omega_{d(m)}$, the quality factor of both resonances experiences a rapid increase shown in Figures 2d,e by the red circles. The rise in the Q-factors (QFs) correlates with the decrease of the real (dissipative) part of graphene's surface conductivity $\sigma_r$ shown by the blue circles in Figures 4d,e. Overall, the transition to Pauli blocking regime for high values of carrier density causes considerable change of the QFs for both modes: the $Q's$ of monopole mode and dipole mode changes in the $14.3 < Q < 16.4$ range (15%) and in the $15.0 < Q < 18.2$ range (21%), respectively. Note that the maximum of QF coincides with the minimum of $\sigma_r$ for both modes.

Spectral blue-shifting of both resonances can be clearly observed in Figures 4b,c as the carrier density increases. This experimentally observed behavior is consistent with eq 2. For example, for the dipole mode the spectral shift is $|\Delta\lambda_d| \approx 0.33 \mu m$ ($\Delta\omega_d \approx 62\ cm^{-1}$), or about $|\Delta\lambda_d|/\lambda_d \approx 4.6\%$ of the total bandwidth. Given the Q-factor of $Q_d \approx 18$, we conclude that the $|\Delta\lambda_d|/\lambda_d > 1/2Q$ condition which is necessary for an efficient and deep modulator is satisfied for the dipole resonance of the integrated graphene/metasurface structure. On the other hand, the corresponding spectral shift numbers for the monopole resonance (EIT1) are more modest: $|\Delta\lambda_m| \approx 0.09 \mu m$ ($\Delta\omega_m \approx 23\ cm^{-1}$), or about $|\Delta\lambda_m|/\lambda_m \approx 1.4\%$. The spectral shift of the monopole mode is considerably smaller than that of the dipole mode, which is mainly due to its weaker interaction with graphene (smaller tangential field enhancement $\eta$) as shown in Figure 1e.

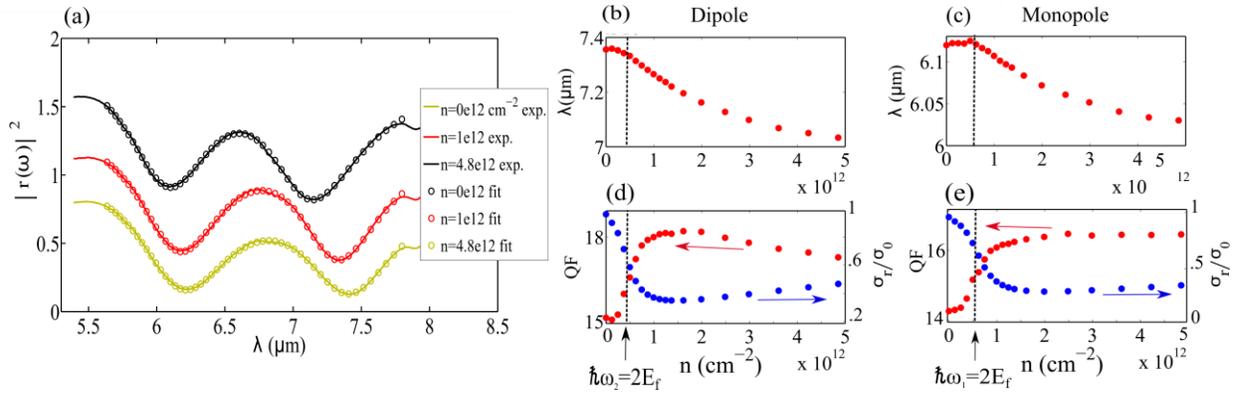

**Figure 4**. **(a)** The spectrum of the experimental reflectivity (solid lines) for the Sample 1, fitted by $|r(\omega)|^2$ from eq 1 (circles) for three selected doping concentrations (holes) $n$ in graphene. **(b,c)** The resonant wavelengths $\lambda_{d,m}(n)$ of the dipole (b) and monopole (c) resonances. **(d,e)** The quality factors (QFs) $Q_{d,m}(n)$ of the dipole (d) and monopole (e) resonances. Dashed line: hole concentration $n$ corresponding to the Fermi level at the Pauli blocking threshold $2E_F(n) = \hbar\omega_{d,m}(n)$. Blue circles: $\sigma_r$ in units of the universal conductance $\sigma_0 = e^2/4\hbar$ at the resonant frequencies $\omega_{d,m}(n)$.

## Comparisons with numerical simulations and prospects for high-speed independent amplitude/phase modulation

Numerical simulations using a commercial finite elements solver COMSOL Multiphysics were carried out in order to verify the experimental results presented in Figures 3,4. In addition, these numerical simulations enable us to make predictions about some of the physical quantities that were not experimentally measured in this work, such as the amplitude and phase of the transmitted waves. The SLG was modeled using a surface current [50] $J_{SLG} = \sigma_{SLG} E_t$. The dielectric permittivities of Au and SiO$_2$ were taken from the Palik handbook of optical constants[71] in all simulations. Figures 5a,b show the simulated amplitudes $R(\lambda) \equiv |r(\lambda)|^2$ and phases $\varphi_R(\lambda)$ of the reflected light for different carrier concentrations, and Figure 5a is in excellent quantitative agreement with the experimentally measured spectra shown in Figure 3a. For completeness, we also present in Figures 5c,d the numerically simulated transmittances $T(\lambda)$ and phases $\varphi_T(\lambda)$ of the transmitted light although no experimental measurements of the transmission were performed at this time. All simulations are performed for P-polarized light with the angle of incidence $\theta_i = 25°$ for consistency with our experimental setup that utilizes a high-NA microscope.

Below we use the results of the numerical simulations to demonstrate the promise of graphene-functionalized Fano-resonant metasurfaces for developing rapidly tunable amplitude and phase modulators (AM and PM) that can potentially operate with nanosecond-scale modulation speeds. We demonstrate that nearly pure phase modulation can be achieved at certain wavelengths (shown by red arrows in Figures 5a,b for reflected light and Figures 5c,d for transmitted light). For other wavelengths (shown by black arrows in Figures 5a,b for reflected light and Figures 5c,d for transmitted light) nearly pure amplitude modulation occurs as the carrier concentration changes. Specifically, in Figure 5e we plot the reflection data for four values of carrier density varying between $n = 1.6e12$ and $n = 4.8e12 \ cm^{-2}$ (color-coded circle symbols) for two specific wavelengths ($\lambda_R^{PM} = 6.7 \mu m$ and $\lambda_R^{AM} = 7.08 \mu m$) in the $(\phi_R, R)$ phase plane. We observe that, for a fixed wavelength $\lambda = \lambda_R^{PM}$, the reflection phase $\varphi_R(\lambda_R^{PM})$ varies by almost $\Delta \varphi_R \approx 20°$ degrees as the function of $n$ while the reflection amplitude $R(\lambda_R^{PM})$ changes by only 3% for the same variation of carrier density. The implication of this result is that one can develop a narrow-band PM which only affects the phase but not the amplitude of the reflected light. For example, if different elements of the metasurface can be independently controlled, one can envision an active beam steering reflect-array that is based on SLG-functionalized metasurface.

Similarly, a narrow-band AM that affects the amplitude but not the phase of the reflected light can be implemented at $\lambda = \lambda_R^{AM}$: according to Figure 5e, the reflectance $R(\lambda_R^{AM})$ varies by factor 5 (from $R = 0.07$ to $R = 0.35$) whereas the phase $\phi_R(\lambda_R^{AM})$ changes by only $\Delta \varphi_R \approx 16°$ degrees for the same variation of graphene's carrier density. Likewise, in Figure 5f we plot the transmission for two selected wavelengths ($\lambda_T^{PM} = 7.46 \mu m$ and $\lambda_T^{AM} = 7.12 \mu m$) in the $(\phi_T, T)$ phase plane. Phase modulation without amplitude change can be achieved in transmission at $\lambda = \lambda_T^{PM}$: the transmission phase $\varphi_T(\lambda_T^{PM})$ varies by almost $\Delta \varphi_T \approx 16°$ degrees as the function of $n$ while the transmission amplitude $T(\lambda_T^{PM})$ changes by only 3% for the same variation of carrier density. Similarly, amplitude modulation without phase change can be achieved in transmission at $\lambda = \lambda_T^{AM}$: the transmission amplitude $T(\lambda_T^{AM})$ changes by a factor 3 (from $T = 15\%$ to $T = 45\%$) while the transmission phase $\varphi_T(\lambda_T^{AM})$ changes by only $\Delta \varphi_T \approx 7°$ degrees. By controlling both the amplitude and phase of the individual segments of a metasurface, one can now envision actively controlled planar structures capable of forming infrared holograms[72] of almost unlimited complexity operating in either transmission or reflection.

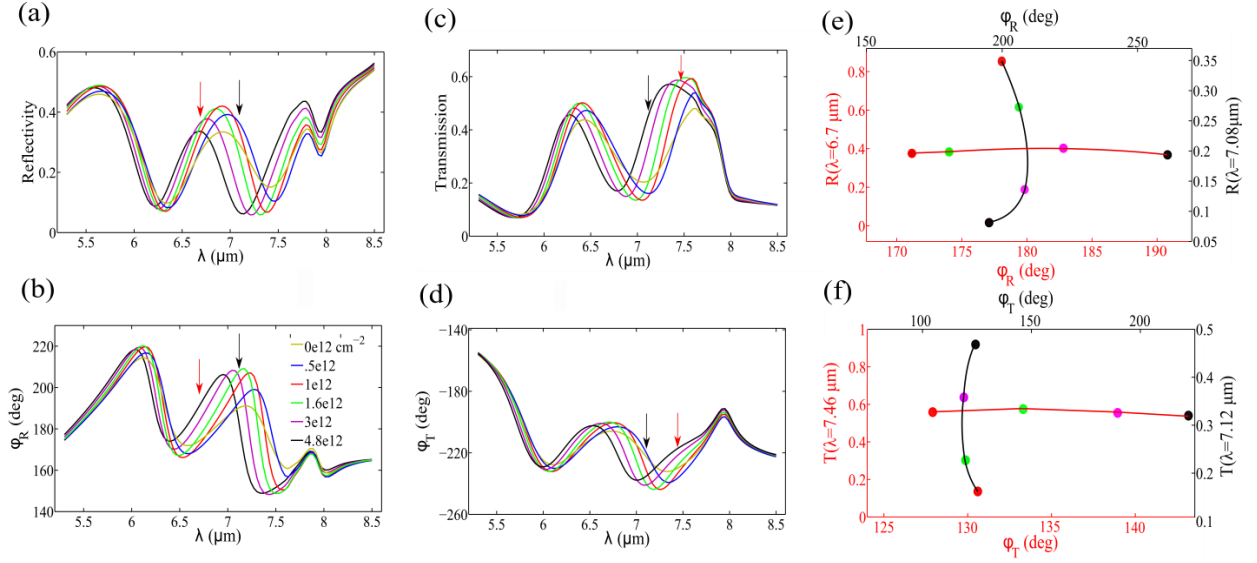

**Figure 5**: Numerical simulations of **(a)** intensities and **(b)** phases of the reflected light, and **(c)** intensities and **(d)** phases of the transmitted light for Sample 1 ($G = 70 nm$) at different graphene doping levels. The arrows indicate the spectral positions with constant scattering intensities (red arrows) and phase (black arrows). **(e)** Differently re-plotted intensities/phase of the reflected light for $\lambda_R^{PM} = 6.7\ \mu m$ (red curves/axes) and $\lambda_R^{AM} = 7.08\ \mu m$ (black curves/axes) for four different doping values. **(f)** Same as **(e)** but for transmitted light at $\lambda_T^{PM} = 7.46\ \mu m$ (red curves/axes) and $\lambda_T^{AM} = 7.12\ \mu m$ (black curves/axes). Graphene doping levels in (e) and (f) are color-coded according to the legend in **(b)**.

The time response of the suggested intensity/phase modulator is limited by the RC time of the device. In estimating the time response of the device we follow the earlier established [55] methodology and evaluate the following capacitances and resistances of the device: the contact pad capacitance $C_P$, graphene capacitance $C_G = C_g A_G$ (where $A_G$ is graphene's area), and circuit resistance $R = R_{MG} + R_G$, where $R_{MG}$ is the resistance between the metal contacts and graphene and $R_G$ is the resistance of graphene. Circuit analysis[55] provides two relevant time constants that limit the time response of the entire device: $\tau_P = 2C_P R_S$, which is related to the voltage source and the connecting pads (where $R_s = 50\Omega$ is the voltage source resistance in our experiment) and $\tau_G = C_G R/2$, which is related to charging graphene itself. Frequency response is limited by the longest of these two time scales. For our sample $A_G = 1e4\ \mu m^2$ and $A_P = 4e4\ \mu m^2$ are the areas of the graphene sheet and the source/drain contacts, respectively. This gives $C_P \approx 3.5 pF$ and $C_G = 0.2 pF$, and according to Figure 2a the circuit resistance as a function of voltage is in the $0.45 k\Omega < R < 4.5 k\Omega$ range. Assuming that $R = 2k\Omega$, we estimate the relevant response times as $\tau_G = 200 ps$ and $\tau_P = 80 ps$. Thus it appears that the charging time of graphene is the limiting factor in determining the modulation speed of the device. The resulting 3dB cut-off frequency based on this estimated value of $\tau_G$ is calculated to be about 3.5 GHz.

**Conclusions:** We have experimentally demonstrated dynamic control of the reflectivity of mid-infrared light using electrically gated single layer graphene. An order of magnitude modulation of the reflected light was accomplished by designing a novel type of a metasurface supporting double Fano resonances and integrating it with an under-layer of graphene. The unique aspect of such modulator is its high baseline reflectivity and large reflectivity extinction coefficient (modulation depth). Numerical simulations indicate that independent amplitude and phase modulation are possible in reflection and transmission. This work paves the way to future development of ultrafast opto-electronic devices such as dynamically reconfigurable holograms, single-detector imagers, dynamical beam-steering devices, and reconfigurable biosensors.

**Acknowledgments:** This work was supported by the Office of Naval Research (ONR) Award N00014-13-1-0837 and by the National Science Foundation (NSF) Award DMR 1120923. I.K., C. M., and R.S.R. would like to acknowledge the support from a Tokyo Electron Ltd (TEL)-customized Semiconductor Research Corporation award (2009-OJ-1873) and the Office of Naval Research (Grant N00014-10-1-0254). Babak Fallahazad, Sommayeh Rahimi, and Nima Asoudegi are acknowledged for their advices on the fabrication procedures.